# Lomonosov's Discovery of Venus Atmosphere in 1761: English Translation of Original Publication with Commentaries.

Vladimir Shiltsev

*FNAL, PO Box 500, Batavia, IL 60510 USA,* e-mail: *shiltsev@fnal.gov*

**Abstract.** Key figure of European Enlightenment, Russian polymath Mikhail Vasilievich Lomonosov (1711—1765) had discovered atmosphere of Venus during planet's transit over the Sun's disc in 1761. This article contains the first complete English translation of his report (originally published in Russian in July of 1761 and in German in August of the same year), commentaries and extensive bibliography.

## PART I: TEXT OF THE ENGLISH TRANSLATION

### "THE APPEARANCE OF VENUS ON THE SUN, OBSERVED AT THE ST.PETERSBURG IMPERIAL ACADEMY OF SCIENCES ON MAY 26, 1761"

1  There is no need to mention at length all the benefits of observing celestial bodies and especially the
2  changes which happen rarely and bring great benefits. Physicists appreciate how useful such thoughtful
3  observations are when studying natural mysteries and educating the human mind. Astronomers know how
4  they can determine the movements of the main bodies of the visible world; geographers know how they can
5  measure and demarcate the globe without error; navigators appreciate the benefits of safe navigation of their
6  ships through the sea.

7  The monarchs and governments having the true care for common good do not spare their resources
8  on constructing astronomical observatories, on supporting and honoring the people who know this very
9  science and undertaking expeditions to remote lands to observe rare celestial phenomena such as the recent
10  appearance of Venus on the Sun, which satisfied the curiosity of many astronomers in European
11  observatories and many in other parts of the world sent from France and England combined with the
12  expansion of useful knowledge. From the local Imperial Academy of Sciences, Court Councilor and
13  Astronomy Professor Popov and Mathematics Adjunct Rumovskii were sent to remote Siberian lands by the
14  highest commandment of H.I.M. Ruling Senate with double salaries and fully satisfied needs in instruments
15  and other supplies. [We hope] they have used all possible diligence in observing this phenomenon and were
16  rewarded with clarity similar to what it was here and gave clear vision to the local observers during all the
17  time of passage of Venus, which appeared on the Sun.

18  In the meantime, while waiting for the information to be submitted to the Academy of Sciences
19  from our and foreign remote observers from different parts of the world, [we] report to the world of scientists
20  our local observations of this rare phenomenon made by Major and Astronomy Adjunct Krasilnikov and Mr.

21 Kurganov who is an assistant in mathematical and navigational sciences at the rank of the lieutenant. To
22 better inform the world of scientists and science-lovers about their extensive skills in astronomy and of their
23 work, a short summary is given below.

24      Mr. Krasilnikov, a student of Professors Delisle and Farquharson, was in charge of astronomical
25 observations in the Kamchatka expedition for 13 years starting in 1733. Upon his return, he went to Narva,
26 Revel, Riga for similar [astronomical] work and to the island of Dago for the exact composition of sea maps.
27 His observations allowed for the determination of the longitudinal expansion of the entire Russian state from
28 the harbor of St. Peter and Paul on the east coast of Kamchatka to Cape Dagerort; he also determined the
29 longitudes and latitudes of many places inside the Russian empire. In 1753 the Academy of Sciences sent
30 him to observe the transit of Mercury across the Sun in Moscow. He completed the task and reported it in the
31 academic Commentaries and Proceedings.

32      Mr. Kurganov practiced astronomy for many years at the Academy's observatory under Mr. Popov
33 and Mr. Krasilnikov. With the former, he was in the aforementioned expedition to Livonia and Estland; then
34 he carried out important astronomical observations on the island of Ezel for over a year with the Astronomy
35 Professor Grischow and was promoted to the Academy's Adjunct by him. In the last year, for his skills in
36 astronomy, he was transferred from the Admiralty Board to the Academy of Sciences to make astronomical
37 observations in order to make corrections in the Russian atlas.

38      Their observations at the local Observatory were arranged as follows. Several days before the
39 appearance of Venus on the Sun, they identified the local noon time with a probable error of less than a
40 second, corresponding to the many heights of the Sun in the mornings and afternoons - as recorded in their
41 logbook - and precisely determined the meridians, so that on the morning of [May] 26 they had true time.
42 Through the six-foot refractor tube with two glasses Mr. Krasilnikov saw:

43          the edge of Venus on the Sun at 4 hours 10'1";
44          full entry of Venus or internal contact by its trailing edge at 4 hours 26'39";
45          at the exit, the first contact of its front edge at 10 hours 19'4";
46          complete egress at 10 hours 37'0".
47 Meantime, Mr. Kurganov saw through the Gregorian telescope:
48          the first edge of Venus on the Sun at 4 hours 9'42";
49          full entry or contact by its trailing edge at 4 hours 26'41";
50          at the exit, the first contact of its front edge at 10 hours 19'1";
51          complete egress at 10 hours 37'2".
52 And inasmuch as their telescopes did not have a functional micrometer, which would have been more
53 convenient for taking important measurements such as the shortest distance from Venus to the Sun's center,
54 its width and other things, then they used another excellent [effective] method to determine its precise path
55 during its [Venus's] passage across the Sun.

56      A parallax machine with a six-foot tube was set up along the meridian line of the Observatory
57 equipped with reticule, that is a mesh of equal silk threads arranged as shown (in figure 8) placed in the tube
58 in such an orientation that the southern edge of the Sun (in reversed image) was always touching one of the
59 silk threads *pe* during daily passage. With all this done, any observation of less than 2 ¼ minutes, the change
60 in the Sun's declination did not matter much, because even the daily difference thereof was less than six
61 minutes. Then in turn one of the observers would watch for the passage of Venus's center through the
62 reticule's threads, while the Sun's edge was touching them, and gave prompt signals, while the other
63 observer, constantly looking at his watch, recorded those moments. The center of Venus in such a passage
64 was well determined, because the passage across its width as a whole took no more than 4 ½ seconds. Nine
65 such observations were taken with a sufficient degree of accuracy, even without a micrometer. As a result,

66 trustworthy calculations were made on the basis of the astronomical theory using the latest solar tables
67 compiled by Mr. de La Caille, producing the following results.

68      The passage of Venus's diameter through the line *cd* has been repeatedly noticed in 4 ½ seconds,
69 and the Sun's [diameter] through the center-point in 2'17". From that, the calculated diameter of the Sun in
70 parts of a larger circle is 0°31'36", while it is 1'2" for Venus. Consequently, the ratio of their diameters is 61
71 to 2. The true [total] time of Venus conjunction with the Sun was 7 hours 43'5". Their length [azimuth] in
72 P[etersburg] was 15°36'9".  Southern latitude of Venus was 0°10'1". The angle of inclination of its path to the
73 eastward latitude circle was 81°29" *.

74 * [footnote] The aforementioned Mr. Kurganov calculated that another memorable passage of Venus across the Sun will
75 happen on May 23, 1769  old style, and though it is doubtful that it can be seen in St. Petersburg; still it could be seen at
76 other places around the local latitude and especially further to the North.  This is because the ingress will start in the 10$^{th}$
77 hour in the afternoon and the egress will occur in  the 3$^{rd}$ hour after midnight, and [Venus] will appear in the upper half
78 of the Sun at a distance close to 2/3 of the solar half-diameter from its center. The next appearance [of Venus on the Sun]
79 will take place one hundred and five years after 1769. Also, on October 29 of the same year, a similar passage of the
80 planet Mercury across the Sun's disc will only be visible in South America.

81      Besides such rigorous  astronomical observations, Mr. Collegiate Councilor and Professor
82 Lomonosov kept a curious look-out at his place mostly for physical observations, using a 4 ½ feet long
83 telescope with two glasses. The tube had attached a not-so-heavily smoked glass, for he intended to observe
84 the beginning and the end of the phenomenon only and then to use the power of the eye, and give [his] eyes a
85 respite for the rest of the transit.

86      Having waited for Venus to appear on the Sun for about forty minutes beyond the time prescribed in
87 the ephemerides, [he] finally saw that the Sun's edge at the expected entry became indistinct and somewhat
88 effaced, although before [it] had been very clear and equable everywhere (see *B*, figure 1); however, not
89 seeing any blackening and thinking that his tired eyes were the cause for this blurring, [he] got away from the
90 tube. After a few seconds, [he] took a glance through the tube and saw that in the place where the Sun's edge
91 had previously appeared somewhat blurred, there was indeed a black pock or segment, which was very small,
92 but no doubt due to the encroaching Venus. Then [he] watched attentively for the entry of the other (trailing)
93 edge of Venus, which seemed to have not yet arrived, and a small segment remained beyond the Sun.
94 However, then suddenly there appeared between the entering trailing edge of Venus and the solar edge, a
95 hair-thin bright radiance separating them, so that the time from the first to the second was no more than one
96 second.

97      During Venus's egress from the Sun, when its front edge was beginning to approach the solar edge
98 and was (just as the naked eye can see) about a tenth of the diameter of Venus, a blister [pimple] appeared at
99 the edge of the Sun (see *A*, fig. 1), which became more pronounced as Venus was moving closer to a
100 complete exit (see fig. 3 and 4). *LS* is the edge of the Sun, *mm* is the Sun bulging in front of Venus. Soon the
101 blister disappeared, and Venus suddenly appeared with no edge (see figure 5); *nn* is a segment, though very
102 small, but distinct.

103      Complete extinction or the last touch of the trailing edge of Venus to the Sun at its very emergence
104 which followed after a small break and was characterized by blurring of the solar edge.

105      While it was happening, it was clearly noticed that as soon as Venus moved off the axis of the tube
106 and approached the proximity of the edges of field of view, a fringe of colors would appear due to the light
107 rays refraction, and its [Venus] edges seemed smeared the farther [it] was from the [tube] axis *X* (fig. 2).
108 Therefore, during the entire observation the tube was permanently directed in such a way that Venus was
109 always in its center, where its [Venus'] edges appeared crispy clear without any colors.

110      From these observations, Mr. Councilor Lomonosov concludes that the planet Venus is surrounded
111 by a significant air atmosphere similar to (if not even greater than) that which surrounds our terrestrial globe.
112 This is because, in the first place, the loss of clearness in the [previously] tidy solar edge $B$ just before the
113 entry of Venus on the solar surface means, as it seems, oncoming of the Venusian atmosphere onto the edge
114 of the Sun. The clarification of this is evident in figure 6. $LS$ - the edge of the Sun, $PP$ is a portion of the
115 Venus's atmosphere. At the time of Venus' egress, the touch of its front edge produced the bulge. It
116 demonstrates nothing but the refraction of solar rays in the atmosphere of Venus. $LP$ is the end of the
117 diameter of the visible solar surface (fig. 7); *sch [jch]* is the body of Venus; *mnn* is its atmosphere; $LO$ is the
118 [light] ray propagating from the very edge of the Sun to the observer's eye tangent to the body of Venus in
119 case of absence of the atmosphere. But when the atmosphere is present, then the ray from the very edge of
120 the Sun $Ld$ is refracted toward the perpendicular at $d$ and reaches $h$, thus, being perpendicularly refracted,
121 arriving at the observer's eye in $O$. It is known from optics that the eye sees along the incident line; thus, the
122 very edge of the Sun $L$, due to the refraction, has to be seen in $R$, along the straight line $OR$, that is beyond
123 the actual solar edge $L$, and therefore the excess of the distance $LR$ should project the blister on the solar edge
124 in front of the leading edge of Venus during its egress.

125 Addition

126      This rarely occurring phenomenon needs two elucidations. The first should be given to allay any
127 unfounded fears and doubts that can exist among people not confined by any doctrine, which sometimes
128 causes disturbances to the general repose. Often the gullible minds listen to and apperceive with horror the
129 prophecies about such celestial phenomena made by wanderers [pilgrims] who not only never heard of
130 astronomy during the course of their whole life but also, walking stooped, simply could not even look up in
131 the sky. Hebetudes of such foolish prophets and gullible listeners should be scorned and ridiculed. And
132 anyone who vexes about such fictitious troubles should be punished for his own stupidity. It all mostly
133 concerns common people who have no idea about sciences. A peasant laughs over the astronomer and
134 considers him to be a person of no great depth. The astronomer keeps sniggering to himself, and, being
135 much superior in knowledge, mentally remembers that [the peasant] was created in his likeness.

136      The second elucidation extends to literate people, to the readers of the scriptures and zealots of
137 Orthodoxy, whose sacred labors are commendable by themselves, though their excesses sometimes hinder
138 the advancement of high sciences.

139      After reading here about the great atmosphere around the aforementioned planet, one can say: we
140 can then presume that because of its vapor updrafts, clouds gather, rains fall, flowing streams gather into the
141 rivers, and the rivers flow into the seas, different types of plants grow everywhere on which animals feed.
142 And all this, one may say, similarly to the Copernican system, contradicts the law.

143      Such kinds of reflections lead to a relevant dispute on the movement and immobility of the Earth.
144 Theologians of the Western Church take the words of Joshua Navin, chapter 10 verse 12, quite literally and
145 therefore want to prove that the Earth is immobile.

146      This dispute was started by pagans, not by Christian teachers. Long before the birth of Christ,
147 ancient astronomers made the following admissions: Nicetas of Syracuse admitted the existence of the daily
148 revolution of the Earth on its axis; Philolaus its yearly rotation around the Sun. A hundred years later,
149 Aristarchus of Samos presented clearly the heliocentric system. However, Hellenic priests and superstitious
150 people opposed these ideas and objected the truth for many centuries. First, someone Cleanthes reported on
151 Aristarchus with his system of the motion of the Earth dared to displace the great goddess Vesta [Hestia], the
152 keeper of all the Earth, dared to constantly turn the Neptune, Pluto, Ceres, all nymphs, gods of the forests and
153 home all over the Earth. Thus, the pagan superstition kept the astronomic Earth in its jaws, not letting it go,

154 while she [the Earth] always dutifully performed her business and God's command. Meanwhile, astronomers
155 were forced to make up stupid cycles and epicycles (circles and secondary circles) for planetary paths which
156 contradicted [celestial] mechanics and geometry.

157 It is a pity that there were no such clever cooks, as the following one:

158 *Once, feasting, two astronomers were seated,*
159 *And argued 'mongst themselves in language heated*
160 *Earth turning travels round the Sun, did one maintain,*
161 *The other that Sun leads the planets in its train.*
162 *This one was Ptolemy, the other one, Copernicus.*
163 *The smiling Cook resolved their quarrel thus:*
164 *Knowst thou the course of stars? the host inquired,*
165 *Then how to solve the question art inspired?*
166 *Copernicus was right, the answer went.*
167 *I'll prove it true, although I've never spent*
168 *Time on the Sun. What Cook of brains could boast*
169 *So few, to turn the Hearth about the Roast?*

170 Finally, Copernicus resumed the solar system which now bears his name and showed its glorious
171 usefulness in astronomy; later Kepler, Newton, and other great mathematicians and astronomers brought it to
172 great accuracy which we now see in predicting celestial phenomena, and which is impossible to achieve in
173 the Earth-centric system.

174 Ineffable wisdom of God's deeds follows from reflections on all the creatures, that's where physics
175 leads [us], but most of all, the astronomy gives us the best understanding of its majesty and power when
176 showing the orderly flow of heavenly bodies. We reveal the Creator more clearly as the precise observations
177 better match our predictions, and as the grasp of new revelations widens, [we] praise Him louder.

178 The Scripture should not always be understood literally, but often rhetorically. St. Basil the Great
179 serves an example of consenting the Scripture with the Nature when in his Discourses on the six days [of
180 creation] he clearly shows how biblical words should be interpreted in appropriate places. Talking about the
181 Earth in general, he writes: "When [you] hear in psalms: "I established her [Earth] foundations" – take the
182 meaningful power of that saying about the pillars" (conversation 1). Discussing the words and
183 commandments of God during the creation, "…And God said," and others, he [St.Basil] declares the
184 following: "the Mighty word gets in through the wit" (conversation 2), clearly expressing that the divine
185 words do require neither the mouth nor ears, nor air to communicate their benevolence, but speak by the
186 power of the mind. And in another place (conversation 3) [he] further confirms in the following discussions:
187 "In Israel's curse, you will have copper skies: What do these words say? Wide-spread drought and
188 diminishing of water in the air." Discussing God's reaction frequently mentioned in the Bible, [he] writes:
189 "And God saw that it was good: Scripture does not tell that a nice view of the sea opened to God. For the
190 eyes of the Creator did not see the beauty of the creatures, but with unspeakable wisdom [he] contemplates
191 the situation. " Seems that this great and holy man has shown enough that elucidation of the sacred books not
192 only is allowed, but sometimes is needed when for the sake of metaphorical expressions there seems to be a
193 contradiction with Nature.

194 Truth and faith are two sisters and daughters of one Almighty Parent: they can never come in a
195 quarrel with one another; unless someone instigates the enmity between them out of vanity and ill sophistry.
196 Prudent and kind people should consider whether there is any way to explain and prevent the imaginary
197 animosity between them, as perpetrated by the aforementioned wise teacher of our Orthodox church.

Following him, St. John of Damascus, profound theologian and holy poet, mentioned different opinions about the structure of the world and said in the "Precise Statement of the Orthodox Faith" (book 2, ch. 6): "Whether it happens this way or that, but all appeared and was approved by the divine commandment". It means that physical explanations of the world are for the glorification of God and not harmful to the faith. Similarly, it is said in the following: "Heavenly heaven is the first heaven, located above the firmament. That makes already two of them, for God called the firmament heaven as well. In the divine Scripture, air is also commonly called the heaven as it is seen above. For it says: bless all the birds of the air, meaning the air as the birds' place, not the heaven. So, here are three heavens mentioned by the divine Apostle. If you wish to understand even seven zones as seven heavens - it does not harm the true reasoning." It means the following: if one accepts ancient Hellenic views about the seven heavens, it is not of harm to the Scripture and the book of Paul.

St. Basil the Great considered the possibility of multiple worlds and said: " As a potter, who made thousands of water pots with the same skill and exhausted no art and no power, the Creator of the universe has enough creative power not only for a single world, but for as many as [He] wants to create in a single moment alone."

That is how these great enlighteners tried to connect the study of nature with the faith by reflections in the books on the base of knowledge of astronomy then. Oh, if in their times today's astronomical instruments had been invented and numerous observations far superior to the ancient astronomers' knowledge of the heavenly bodies had been made, if then thousands of new stars accompanied by new phenomena had been discovered. then with what a high soaring spirit and superb eloquence those holy rhetoricians would have preached the majesty, wisdom and power of God!

Some people ask if there are humans like us living on the other planets, then what faith are they? Have they been preached the gospel? Are they baptized into the faith of Christ? To those [we] give the answer by question. In the Southern great lands, the coasts of which have just recently been discovered by modern seafarers, local inhabitants, as well as the inhabitants of other unknown lands, the people of a different type, language and behavior than us - what faith are they in? And who preached the gospel to them? If one wants to know that or wants to convert and baptize them, let him go with the Gospel words ("Acquire neither gold nor silver [nor brass in your belts, don't stop for feasts on the way, and care for no robe, no shoes, no wand]"). And as his sermon is finished, then let [him] go to do same on Venus. If only his labors had been not in vain. Maybe the [Venusian] people did not sin in Adam, and all the consequences are of no need. "There are many paths to salvation. There are many places in heavens. "

With all this, the Christian faith is immutable. It cannot be repugnant to God's creation, and the only contradiction is such that does not delve in the creations of God.

The Creator has given two books to the mankind. In one [He] has shown His majesty, in another - His will. The first one is this visible world, established by Him so that a man looking at the vastness, beauty and elegance of its buildings, acknowledges divine omnipotence, as much as he can understand. The second book is the Holy Scripture. It shows His favor to our salvation. In these prophetic and apostolic God-inspired books, the interpreters are our great teachers of the Church. In the book of creation of the visible world, the interpreters are physicists, mathematicians, astronomers and other expounders of the divine infusions into the nature - much like the prophets, apostles and teachers of the church in the scripture. Unreasonable is mathematician, if he wants to measure the divine will by compass. The same is true about the theology teacher if he thinks that astronomy or chemistry can be learned from the Psalms.

Commentators and preachers of the Scripture show the way to virtue, reward of the righteous, punishment of law-breakers and preaching well-being accordant with the will of God. Astronomers discover

242 the temple of the divine power and glory and seek the ways to our transitory bliss combined with reverence
243 and thanksgiving to the Almighty. Together both convince us not only in the existence of God but also in His
244 ineffable goodness to us. It is a sin to plant weeds and strives between them!

245       Not only the Hellenic poets, but also the first great Christian teachers give us examples of how
246 understandings and observations of the natural things do strengthen [us] in the faith.

247       In [the book]"Against Rufinus", Claudian announces  how helpful studies of Nature are for the
248 knowledge of God:

249 *I have been long and long in thoughtful doubt,*
250 *Is there oversight to Earth from Heights,*
251 *Or things just blindly flow without order,*
252 *And Heavens do not run the universe.*
253 *But seeing harmony in heaven's bodies,*
254 *The order in the Earth, and seas and rivers,*
255 *Appearance of the Moon and change of days and nights,*
256 *I do acknowledge - we are God's creations.*

257       There is nothing else left to add, but to summarize and repeat that understanding of the nature, in its
258 any manifestation, is not against the Christian law, and he who tries to explore the nature, knows and honors
259 God and  agrees with the words of Basil the Great (conversation 6, on the existence of stars): "In this study,
260 we come to know ourselves,  understand God, worship the Creator, assist  the Lord, praise the Father,  love
261 our Feeder, honor the Benefactor, and continue to worship the Architect of our present and future life - the
262 One whose bestowed wealth certifies the promised benefits, and [Who] makes anticipations undoubted as we
263 explore the present."

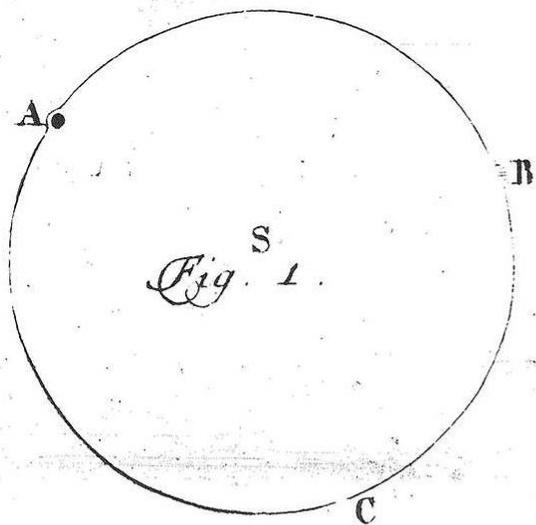

A

B

*Fig. 1.* S

C

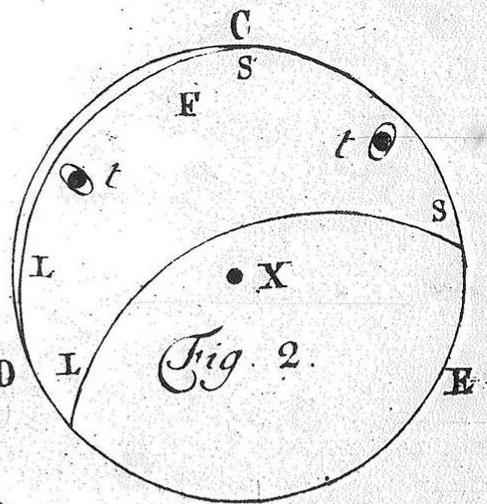

C
S
F
t        t
L                    s
L        X
D    L        *Fig. 2.*        E

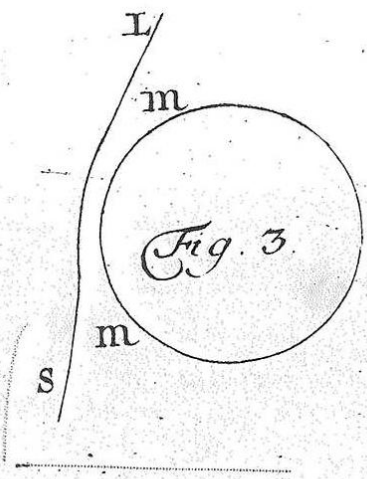

L
m

*Fig. 3.*

m
S

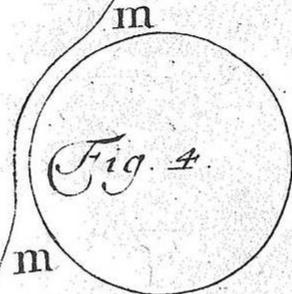

m

*Fig. 4.*

m
s
L

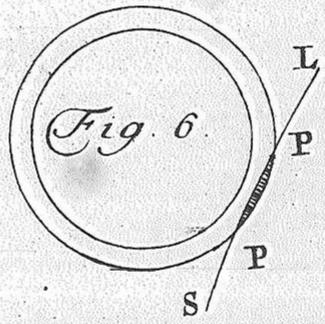

L
*Fig. 6.*    P

P
S

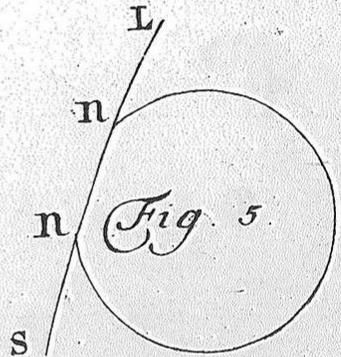

L
n

*Fig. 5.*

n
S

*Fig. 8.*

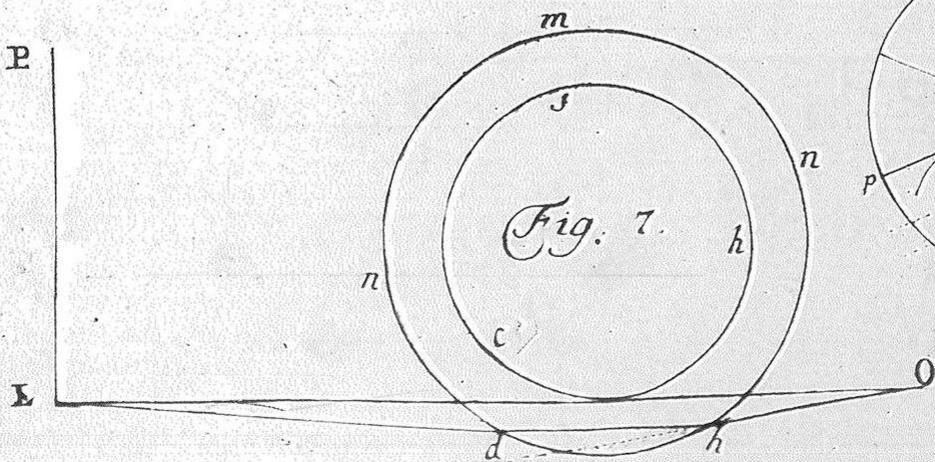

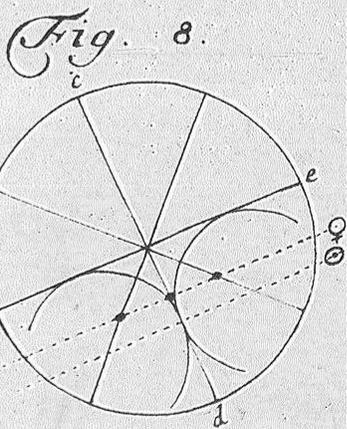

P

m
f

n                    n

*Fig. 7.*    h

n

c

x

d        h        O

R

# PART II: COMMENTARIES

Russian polymath Mikhail Vasilievich Lomonosov (1711—1765) – see more on him , e.g., in (Shiltsev, 2012) and references therein - had discovered atmosphere of Venus during planet's transit over the Sun's disc in 1761. The text presented above is the first complete translation of M.V.Lomonosov's paper (Lomonosov, 1761a) from Russian to English. It has been done by Vladimir Shiltsev in consultation with Natalia Eidelman (Novosibirsk State University, Chair of English Language, Novosibirsk, Russia) and Dr.Igor Nesterenko (Budker Institute of Nuclear Physics, Novosibirsk, Russia). Several comments and corrections by Randall Rosenfeld (Canada) and Yuri Petrunin (Telescope Engineering Company, CO, USA) are greatly acknowledged, too. The text in brackets […] is added by the translator, who tried to provide word-by-word translation where possible and stay as close as possible to the original. The original text in Russian is taken from (*Complete Works*, 1955).

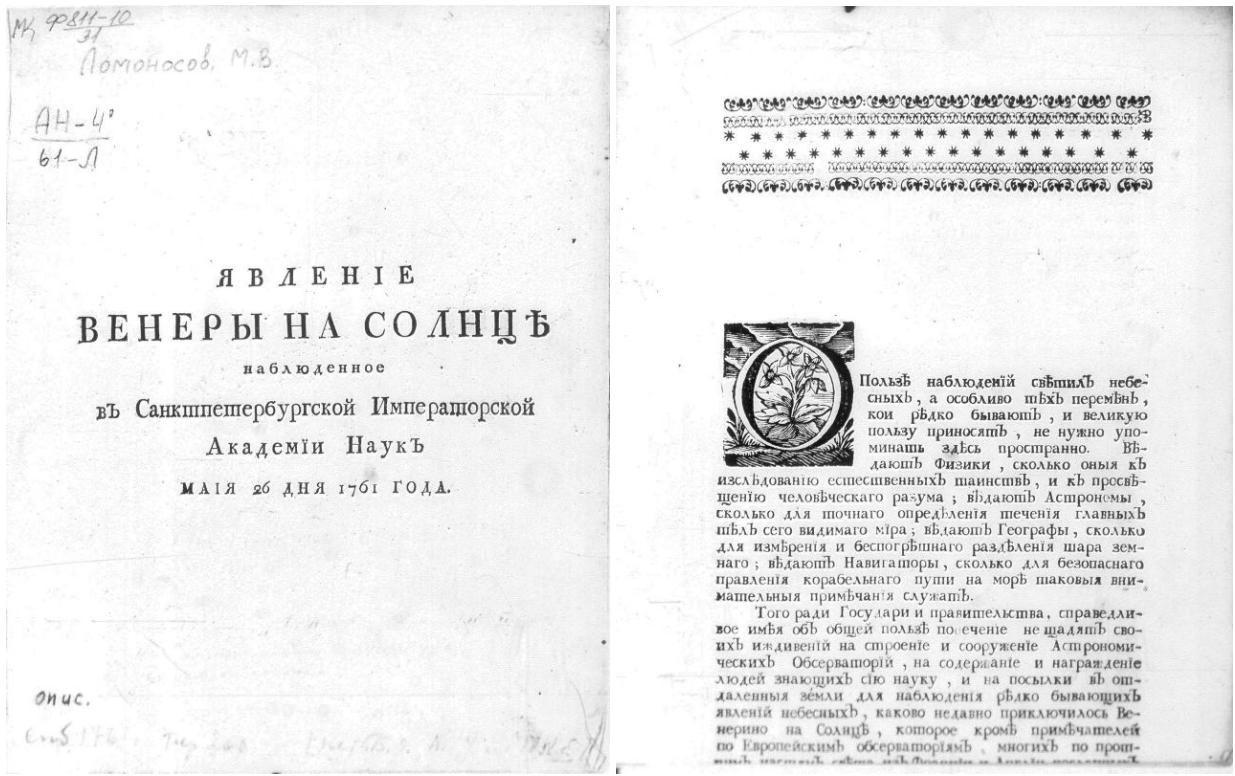

Figure I: The title page and the first page of the original Russian 1761 publication of "The Appearance of Venus …" (submitted July 4, published July 17, 1761).

Mikhail Lomonosov had submitted his article "The Appearance Of Venus On The Sun, Observed at the St.Petersburg Imperial Academy Of Sciences On May 26, 1761" (in Russian) for publication on July 4[th], 1761 (all dates in these commentaries are in old style), and 250 copies were published by the St.Petersburg Imperial Academy of Sciences Print on July 17, 1761 – according to the records of the Russian National Library. Over the next few months, the entire Russian edition copies were sold: 50 copies were distributed free of charge, two ended in the Academy's library, six given to the author, 147 sold in the St. Petersburg bookstore, 30 were sent for sale to Moscow, 5 were stored in the Academy's bookstore as unliable for sale (Tyulichev, 1988). The German translation (Lomonosov, 1761b) was made shortly (presumably by Lomonosov himself [5]) and 250 copies were printed in August 1761 and sent for wide distribution abroad (Sharonov, 1952a; Chenakal and Sharonov, 1955;Tyulichev, 1988). The Russian and German texts differ by eight insignificant textual words/phrases (Chenakal and Sharonov, 1955). The title page and the 1[st] page of the Russian original article are presented in Figure I; the title page and the 1st page of the German translation from the collection of digitized books of the Universitäts- und Landesbibliothek Sachsen-Anhalt (originally from the University of Halle library) are shown in Figure II.

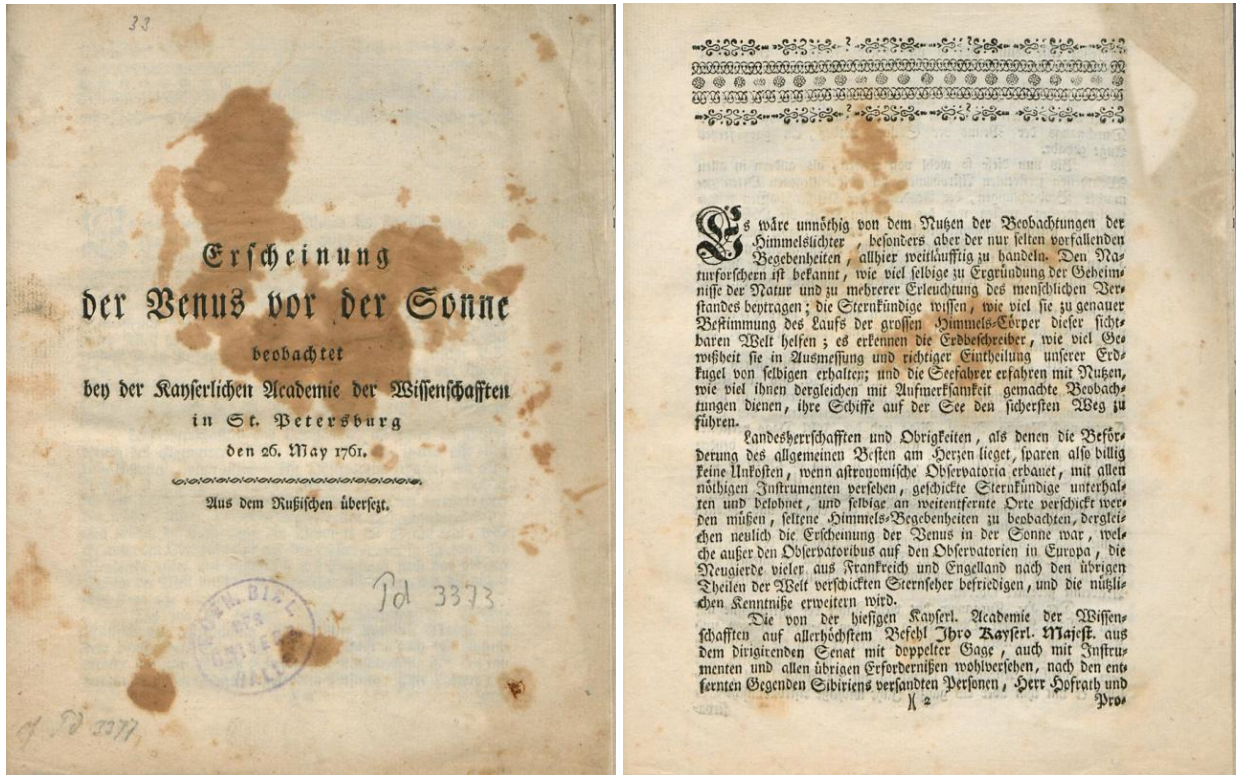

Figure II: The title page and the first page of the 1761 publication of Lomonosov's "Erscheinung der Venus vor der Sonne beobachtet …" in German (published in August, 1761)

The paper has appeared in all major editions of Lomonosov's *Complete Works* published by St.Petersburg-USSR-Russian Academy of Sciences (e.g., *1803*, Vol.3, p.260; *1950-1983*, Vol.4, p.361; *2011*, Vol.4, p.219). The latter two also contain five related notes, articles and preparatory materials and are well commented. The article itself, its physics content and importance have been discussed in great detail in several publications by leading Russian astronomers (Sharonov, 1952a; Sharonov, 1952b; Sharonov, 1955; Sharonov, 1960; Melnikov, 1970 -all in Russian) and (Marov, 2004).

The commentaries below follow the line numeration of the above posted text of the English translation.

1        *…all the benefits of observing celestial bodies…*        Lomonosov's own first publication on astronomy is dated to 1744, i.e., 17 years before the Transit of Venus in 1761 - see article #1 in (*Complete Works,* 1955)

7        *…constructing astronomical observatories …*        E.g., the state-supported St.Petersburg Imperial Academy of Sciences had built a world-class Observatory by 1727, which was consequently led by such prominent astronomers as Profs. (Academicians) J.N.Delisle (also - De l'Isle, from 1727 until 1747), C.N.Winsheim (1747-1751), A.N.Grischow (1751-1760) - see detail discussion in (Chenakal, 1955; Chenakal, 1957). F.U.T Epinus (also – Aepinus in Latin publications) was the Director of the Observatory in 1761.

12        *… Court Councilor and Astronomy Professor Popov and Mathematics Adjunct Rumovskii …* Nikita Ivanovich Popov (1720-1782) and Stepan Yakovlevich Rumovsky (1734-1812) successfully performed the

1761 transit of Venus observations in Irkutsk and in Selenginsk, correspondingly. M.V.Lomonosov, as a member of the Academy's triumvirate Chancellery (as a "chief scientist") was a key organizer of their expeditions and the entire program of the observations in Russia (Chenakal and Sharonov, 1955; Sharonov, 1960) which also included Lomonosov himself (at his home observatory – see below), AD.Krasilnikov and N.G.Kurganov (in Academy's Observatory), J.A.Braun, F.U.T.Epinus (both at private observatories) observing in St.Petersburg and French astronomer J. Chappe d'Auteroch who was invited by the Academy to observe the transit in Tobolsk. So, the total number of observers in Russia was 8.

20      *... Major and Astronomy Adjunct Krasilnikov and Mr. Kurganov ...*      By 1761, Andrei Dmitrievich Krasilnikov (1705-1773) had 29 years of experience in astronomical and geodetic observations. Biography of A.D.Krasilnikov is given in (Nevskaya, 1957). Nikolai Gavrilovich Kurganov (1726-1796) was his apprentice.

24      *... Farquharson ...*      H.Farquharson (Fargwarson, 1674-1739), Scot mathematician and astronomer in Russian service since 1700.

26      *...Revel, Riga for similar [astronomical] work and to the island of Dago…*      Revel is now Tallinn, capital of Estonia, Dago is now Hiiumaa, the second largest island in Estonia.

28      *...Cape Dagerort...*      the most Western point of Hiiumaa island.

33      *...Livonia and Estland…* now – Lithuania and Estonia.

34      *…island of Ezel…*      now - Saaremaa, the largest island in Estonia.

38      *…local Observatory…*      the St.Petersburg Academy's Observatory, nowadays' address Universitetskaya Embankment, 3, St.Petersburg, Russia. Originally, F.U.T.Epinus planned to observe the transit of Venus from the Observatory. Nevertheless, Epinus's preparations, analysis and predicted times and circumstances of the transit were qualified by Lomonosov as unsatisfactory and Epinus was ordered to allow Krasilnikov and Kurganov to perform the observations at the Academy's Observatory together with him. Epinus refused and went for observations somewhere else. The story of the altercations between Lomonosov and Epinus on this subject is described in detail in (Chenakal and Sharonov, 1955) and commentaries to other articles published in Vol.4 of the Lomonosov's *Complete Works* (1955), as well as in (Marov, 2004 - in English).

54      *…they used another excellent  method to determine its precise path during its [Venus's] passage across the Sun…*      the method presented below over lines 56-64, was also used by A.D.Krasilnikov during the transit of Mercury observations in Moscow on April 23, 1753 (Nevskaya,1957).

66      *... the latest solar tables compiled Mr. de La Caille...* the cited book is *Tabulae solares, quaz e novissimus suis observationibus deduxit N.L.de La Caille, Paris, 1758;* N.L. de La Caille (1713-1762), French astronomer, member of l'Académie des Sciences, one of the organizers of the international campaign of observation of the transit of Venus in 1761; per Lomonosov's *Chronicles* (Chenakal, *et. al,* 1961) "…May 26, 1760. [Lomonosov] Attended the Academic Assembly, where La Caille's letter was read which provided details of the plan of the May 26, 1761 transit of Venus observations by French astronomers."

82      *...at his place...*      the observation had been performed at the Lomonosov's own house in St.Petersburg (modern address Bolshaya Morskaya, 61 – see Figure III), latitude 59°55′50″ N, longitude 30°17′59″E, some 1.3 km South of the Academy's Observatory – see map on Figure IV. Per (Chenakal,

1957) the home observatory was on the flat open roof of a 6m (l) by 5m (w) by 4m (h) building, equipped with some ¾ m high handrails  (the estate destroyed during reconstruction in mid-XIX century).

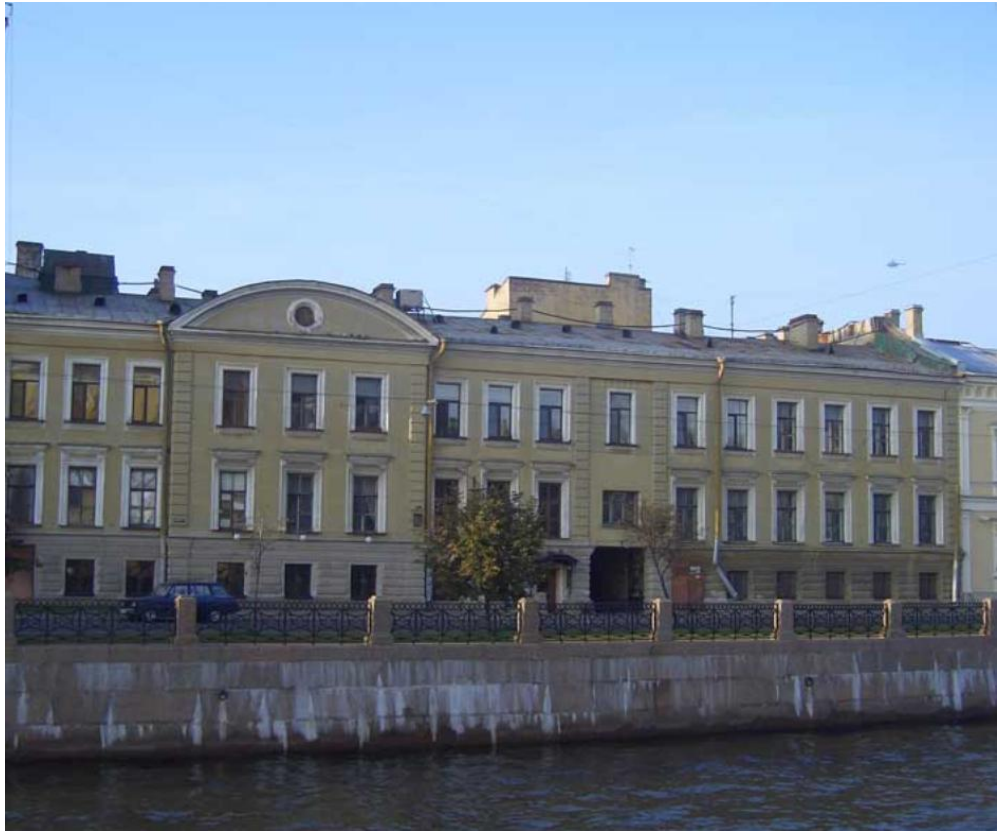

Figure III: Modern days look of the house at Bolshaya Morskaya, 61 in St.Petersburg on the embankment of river Moyka. Lomonosov's original house was reconstructed in the mid-XIX century.

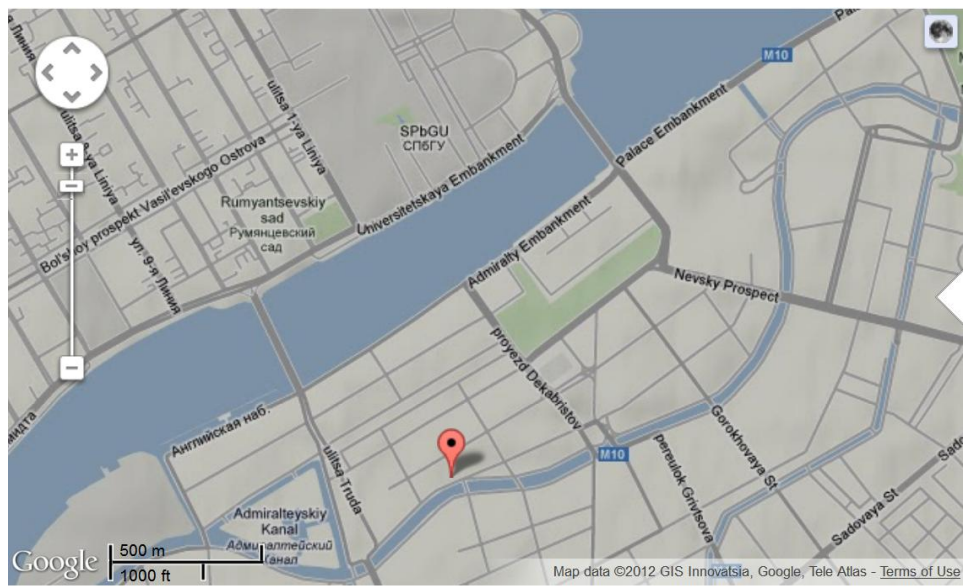

Figure IV: Map of St.Peterburg, which indicates location of the 1761 Lomonosov's house.

82    *...a 4 ½ feet long telescope with two glasses...*    the track of the actual telescope used by Lomonosov has been lost. From the text itself and Figures one can conclude that Lomonosov used an astronomical refractor telescope (with reversed image, therefore, it definitely was not "a sort of spyglass" as incorrectly translated in [11]) with a two-lens achromatic objective; indirect evidences presented in (Melnikov, 1977; Petrunin, 2012) suggest that it possibly was one of the early two-lens achromat refractors made by John Dollond (English optician, 1706-1761). A direct evidence that Lomonosov used the Dollond achromat is given in pre- World War II publication (Nemiro, 1939) .

83    *...not-so-heavily smoked glass...*    in the Russian original - "весьма негусто копченое стекло / ves'ma negusto kopchenoe steklo)", means that Lomonosov used a very weak solar filter; further in the text he notes the need to give regular rest to his eyes.

86-104   *...Having waited for Venus to enter on the Sun (etc)...*    to summarize what Lomonosov had observed: "blurriness" of the Sun's edge at the time of the 1$^{st}$ contact and the 4$^{th}$ contact (illustrated in his Fig.1 at point *B*), a "hair thin bright radiance" close to the 2$^{nd}$ contact which lasted about a second (not illustrated), and "blister" or "bulge" (in Russian – *пупырь/pupyr'* and *выпуклость/vipuklost'*) which lasted for a few minutes after contact 3 (see next comment, illustrated in Lomonosov's Figs. 3,4,5 and at the point *A* in Fig.1).

101    *...blister disappeared, and Venus suddenly appeared with no edge (see figure 5)...*    Figs.3-5 indicate that the "blister" or "bulge" at the third contact appeared from beginning of the egress (the egress phase 1.0), when Venus was fully on the Sun's disc, to the phase of about 0.9-0.94. Therefore, the effect was observed by Lomonosov for one to two minutes.

110-124    *...From these observations, Mr. Councilor Lomonosov concludes that the planet Venus is surrounded by a significant air atmosphere...*    Lomonosov concludes the existence of the atmosphere on base of 3 out of the 4 phenomena observed by him (see comment to lines 86-104 above): "blurriness" of the Sun's edge at the time of the 1$^{st}$ contact and the 4$^{th}$ contact (his physical reasoning illustrated by Fig.6), and the "blister" or "bulge" after the contact 3 (his correct physical explanation was illustrated by Fig.7). V.Sharonov in (Sharonov, 1955; Chenakal and Sharonov, 1955) argued that the "hair thin bright radiance" close to the 2$^{nd}$ contact might also be a manifestation of the refraction of Sun rays in the atmosphere of Venus. In (Sharonov, 1955; Chenakal and Sharonov, 1955; Sharonov, 1960) V.Sharonov has undertaken a detailed comparison of Lomonosov's observation with reports of other observers of the 1761 transit of Venus, which noted some of the above mentioned effects caused by the Venusian atmosphere – e.g., S.Rumovsky, J. Chappe d'Auteroch, T.Bergman and P.Wargentin (effects at ingress/egress) and S.Dunn and B.Ferner (aureole around Venus while on the Sun's disc). On the grounds of: a) the time of the publication of the scientific report; b) completeness of the observations and detailed description; and c) being the only one who has given a correct physical explanation of the effect – Lomonosov's priority was undoubtedly established. Comparison of Lomonosov's 1761 results with observations of the Venus atmosphere effects during the transits in 1769, 1874, 1882, 2004 and 2012 will be a subject of a separate analysis and publication.

144    *…Joshua Navin…*    Joshua was one of Israel leaders, Moses' assistant, and a central character in *Book of Joshua*, the sixth book in the Hebrew Bible and of the Old Testament, cited by Lomonosov.

147    *…Nicetas of Syracuse…*    Nicetas of Syracuse (approximately 400 BC), ancient Greek astronomer.

148     …*Philolaus*…            Philolaus, (c.470–c.385 BC), a pupil of Pythagoras, ancient Greek philosopher, mathematician and  astronomer.

149     ….*Aristarchus of Samos*…            Aristarchus (310 BC – ca. 230 BC), an ancient Greek astronomer and mathematician who presented the first known model that placed the Sun at the center of the known universe with the Earth revolving around it.

150     …*Cleanthes*…     Cleanthes of Assos (c. 330 BC – c. 230 BC), Greeks stoic philosopher, the successor to Zeno as the second head of the Stoic school in Athens.

178     ...*St. Basil the Great*...     Basil of Caesarea, also called Saint Basil the Great, (329 or 330 – 379) Greek bishop of Caesarea Mazaca in Cappadocia, Asia Minor (modern-day Turkey), an influential Orthodox theologian; his Lenten lectures on the Hexaёmeron (on the Six Days of Creation) are cited by Lomonosov here, at the line 209 and at the very end of the article, line 259.

198     …*St.John of Damascus*… Saint John of Damascus, also known as John Damascene (c. 645 or 676 – 749), Syrian monk and priest,  also a polymath whose fields of interest and contribution included law, theology, philosophy and music.

224     ..."*Acquire neither gold nor silver [nor brass in your belts, don't stop for  feasts on the way, and care for no robe, no shoes, no wand]*")...     Matthew, 10:9.

247     ...*Claudian*...     Claudius Claudianus, usually known as Claudian (c.370 –404 AD), Latin poet at the court of the emperor Honorius at Mediolanum (Milan).